# The crystallographic quaternions and their product law

**Cyril Cayron**[a]

[a] Ecole Polytechnique Fédérale de Lausanne, EPFL-STI-LMTM, rue de la Maladière 70, 2000 Neuchâtel, Switzerland

**Synopsis** Unit quaternions are often used to encode rotations in Cartesian bases. Here we how they can be generalized to non-Cartesian bases.

**Abstract** Unit quaternions are widely used in science to encode rotations because the quaternion product is more efficient than matrix product and more stable than Rodrigues product to calculate the composition of two rotations. However, quaternions in their usual form refers to a Cartesian basis; they cannot be used in crystallography as they are. The usual way to solve this issue to apply back-and-forth coordinate changes from the crystal basis to a Cartesian basis attached to the crystal with the help of the structure tensor. Here, we show that actually quaternions can be used directly in the crystal basis by generalizing the quaternion product law. In that aim, we introduced a matrix that we called "cross tensor". It allows the calculation of the cross product in the crystal basis, a bit like the metric tensor allows it for the scalar product. We also show the cross tensor is proportional to the inverse of the metric tensor. The formula of crystallographic quaternion product is then given; it depends uniquely on the metric tensor. The application of the crystallographic quaternions to Electron Back Scatter Diffraction is discussed.

## 1. Introduction

### 1.1. Cartesian rotational quaternions

A rotation $\mathcal{R}$ in three dimensions can be represented by a unit quaternion (Hamilton, 1847). For a rotation of angle α around the axis $\mathbf{u}$, the associated rotational quaternion is

$$\mathbf{q} = \mathrm{c} + \mathrm{s}\,\tilde{\mathbf{u}} \tag{1}$$

with c and s the cosine and sine of the semi-angle of the rotation, $\mathrm{c} = \cos(\frac{\alpha}{2})$, and $\mathrm{s} = \sin(\frac{\alpha}{2})$, and the tilde symbol over $\mathbf{u}$ means that this vector is normalized, i.e. the norm of $\tilde{\mathbf{u}}$ in the Cartesian basis is $\sqrt{\tilde{\mathbf{u}} \cdot \tilde{\mathbf{u}}} = \sqrt{\tilde{\mathbf{u}}^{\mathrm{t}}\,\tilde{\mathbf{u}}} = 1$.

If the rotations $\mathcal{R}_1$ and $\mathcal{R}_2$ are represented by the quaternions $\mathbf{q}_1$ and $\mathbf{q}_2$, respectively, their composition is given by the quaternion product

$$\mathbf{q}_3 = \mathbf{q}_1\mathbf{q}_2 = \mathrm{c}_3 + \mathrm{s}_3\,\widetilde{\mathbf{u}_3} \quad \text{with} \tag{2}$$

$$\mathrm{c}_3 = \mathrm{c}_1\mathrm{c}_2 - \mathrm{s}_1\mathrm{s}_2\,\widetilde{\mathbf{u}_1}.\widetilde{\mathbf{u}_2} \tag{3}$$

$$\mathrm{s}_3\,\widetilde{\mathbf{u}_3} = \mathrm{c}_2\,\mathrm{s}_1\,\widetilde{\mathbf{u}_1} + \mathrm{c}_1\mathrm{s}_2\,\widetilde{\mathbf{u}_2} + \mathrm{s}_1\mathrm{s}_2\,\widetilde{\mathbf{u}_1}\times\widetilde{\mathbf{u}_2} \tag{4}$$

This composition law (2)-(4) is usually proved by using the primitive expression of unit quaternion introduced by Hamilton $\mathbf{q} = a + b\,\mathbf{i} + c\,\mathbf{j} + d\,\mathbf{k}$, with $a, b, c, d$ real numbers such that $a^2 + b^2 + c^2 + d^2 = 1$, and $\mathbf{i}, \mathbf{j}, \mathbf{k}$ pure imaginary numbers that follow the rules $\mathbf{i}^2 = \mathbf{j}^2 = \mathbf{k}^2 = \mathbf{i}\,\mathbf{j}\,\mathbf{k} = -1$ and $\mathbf{i}\,\mathbf{j} = \mathbf{k}$, $\mathbf{j}\,\mathbf{k} = \mathbf{i}$ , $\mathbf{k}\,\mathbf{i} = \mathbf{j}$. However, since in this work the rotational quaternions were directly defined by equation (1) independently of the imaginary numbers $\mathbf{i}, \mathbf{j}, \mathbf{k}$, we propose in Appendix 1, a demonstration based on the spherical laws of angles. These laws themselves are proved in Appendix 2 by using the metric tensor associated to the lattice formed by the three rotation axes.

Note : As for matrices, the meaning of product depends on the active or passive meaning of the quaternions. If the rotations $\mathcal{R}_1$ and $\mathcal{R}_2$ are both written in the same Cartesian basis $(\tilde{\boldsymbol{x}}, \tilde{\boldsymbol{y}}, \tilde{\boldsymbol{z}})$ by the quaternions $\mathbf{q}_1$ and $\mathbf{q}_2$, respectively, then $\mathbf{q}_1\mathbf{q}_2$ is the quaternion given in $(\tilde{\boldsymbol{x}}, \tilde{\boldsymbol{y}}, \tilde{\boldsymbol{z}})$ that represents the rotation that results from $\mathcal{R}_2$ (1st action) followed by $\mathcal{R}_1$ (2d action). The product is right-to-left as for functions; it is also called "active" because $\mathcal{R}_1\mathcal{R}_2$ transforms the vector $\boldsymbol{v}$ given in $(\tilde{\boldsymbol{x}}, \tilde{\boldsymbol{y}}, \tilde{\boldsymbol{z}})$ into a new vector $\boldsymbol{v}' = \mathcal{R}_1\mathcal{R}_2.\boldsymbol{v}$ , also given in $(\tilde{\boldsymbol{x}}, \tilde{\boldsymbol{y}}, \tilde{\boldsymbol{z}})$. If the first step is a rotation $\mathcal{R}_1$written in the Cartesian basis $(\tilde{\boldsymbol{x}}, \tilde{\boldsymbol{y}}, \tilde{\boldsymbol{z}})$ by the quaternion $\mathbf{q}_1$, and that this rotation transforms the basis $(\tilde{\boldsymbol{x}}, \tilde{\boldsymbol{y}}, \tilde{\boldsymbol{z}})$ into the new basis $(\widetilde{\boldsymbol{x}'}, \widetilde{\boldsymbol{y}'}, \widetilde{\boldsymbol{z}'})$, and the second step is a rotation $\mathcal{R}_2$ written in $(\widetilde{\boldsymbol{x}'}, \widetilde{\boldsymbol{y}'}, \widetilde{\boldsymbol{z}'})$ by the quaternion $\mathbf{q}_2$, then $\mathbf{q}_1\mathbf{q}_2$ is the quaternion given in $(\tilde{\boldsymbol{x}}, \tilde{\boldsymbol{y}}, \tilde{\boldsymbol{z}})$ that represents the rotation that results from $\mathcal{R}_1$ (1st operation) followed by $\mathcal{R}_2$ (2d operation). Indeed, $\mathcal{R}_1$ and $\mathcal{R}_2$were considered as coordinate transformation operations $[(\tilde{\boldsymbol{x}}, \tilde{\boldsymbol{y}}, \tilde{\boldsymbol{z}}) \to (\widetilde{\boldsymbol{x}'}, \widetilde{\boldsymbol{y}'}, \widetilde{\boldsymbol{z}'})]$ and $[(\widetilde{\boldsymbol{x}'}, \widetilde{\boldsymbol{y}'}, \widetilde{\boldsymbol{z}'}) \to (\widetilde{\boldsymbol{x}''}, \widetilde{\boldsymbol{y}''}, \widetilde{\boldsymbol{z}''})]$ , respectively, and $\mathcal{R}_1\mathcal{R}_2$ as the coordinate transformation operation $[(\tilde{\boldsymbol{x}}, \tilde{\boldsymbol{y}}, \tilde{\boldsymbol{z}}) \to (\widetilde{\boldsymbol{x}''}, \widetilde{\boldsymbol{y}''}, \widetilde{\boldsymbol{z}''})]$. This product is left-to-right, and is called "passive". The coordinates of a fixed vector $\boldsymbol{v}$ are changed by $\boldsymbol{v}_{/(\tilde{x},\tilde{y},\tilde{z})} = \mathcal{R}_1\mathcal{R}_2\,\boldsymbol{v}_{/(\widetilde{x''},\widetilde{y''},\widetilde{z''})}$.

Rotational quaternions can also be written compactly by $\mathbf{q} = \mathrm{c} + \mathbf{u}$ with $\mathbf{u} = \mathrm{s}\,\tilde{\mathbf{u}}$. The product of two quaternions $\mathbf{q}_1 = \mathrm{c}_1 + \mathbf{u}_1$ and $\mathbf{q}_2 = \mathrm{c}_2 + \mathbf{u}_2$ is then written by

$$\mathbf{q}_3 = \mathbf{q}_1\mathbf{q}_2 = \mathrm{c}_3 + \mathbf{u}_3 \quad \text{with} \tag{5}$$

$$\mathrm{c}_3 = \mathrm{c}_1\mathrm{c}_2 - \mathbf{u}_1.\mathbf{u}_2 \tag{6}$$

$$\mathbf{u}_3 = \mathrm{c}_2\,\mathbf{u}_1 + \mathrm{c}_1\mathbf{u}_2 + \mathbf{u}_1\times\mathbf{u}_2 \tag{7}$$

Since only three free parameters are required for a 3D rotation, even more compact forms exist, such as the triplets of Euler angles, generally associated with rotation matrices, and the Rodrigues vectors, derived from quaternions. A Rodrigues vector $\mathbf{r}$ inherited from a rotational quaternion $\mathbf{q} = \mathrm{c} + \mathrm{s}\,\tilde{\mathbf{u}}$ is simply given by the vectorial part of the quaternion $\mathbf{r} = \left(\frac{1}{c}\right)\mathbf{q} = 1 + \mathrm{t}\,\tilde{\mathbf{u}}$ with $\mathrm{t} = \tan(\frac{\alpha}{2})$; it is the vector $\mathbf{r} = \mathrm{t}\,\tilde{\mathbf{u}}$. The product of two Rodrigues vectors $\mathbf{r}_1 = \mathrm{t}_1\,\widetilde{\mathbf{u}_1}$ and $\mathbf{r}_2 = \mathrm{t}_2\,\widetilde{\mathbf{u}_2}$ is (Morawiec & Field, 1996)

$$\mathbf{r}_3 = \mathbf{r}_1\mathbf{r}_2 = \mathrm{t}_3\,\widetilde{\mathbf{u}_3} = \left(\frac{1}{1-\mathbf{r}_1.\mathbf{r}_2}\right)(\mathbf{r}_1 + \mathbf{r}_2 + \mathbf{r}_1\times\mathbf{r}_2) \tag{8}$$

Rodrigues vectors are compact and they are very helpful to represent the rotations in a Cartesian 3D space that has the same symmetries as the crystal. For example, the corners of a such space correspond to the rotations with the maximum rotation angle; the calculations show that they are 62.80°, 93.84°, 98.42°, 56.60° for cubic, hexagonal, tetragonal and orthorhombic crystals, respectively (Heinz & Neumann, 1991). The advantages of Rodrigues vectors are however outweighed by some singularity issues that are difficult to avoid. Rodrigues vectors corresponding to half turn rotations are indeed infinite vectors, and such problematic vectors can emerge even when two non-infinite vectors are composed. For example, let us consider two rotations of + 90°, $\mathcal{R}_1$ around the x-axis, and $\mathcal{R}_2$ around the z-axis. Their quaternions are $\mathbf{q}_1 = \frac{1}{\sqrt{2}} + \left[\frac{1}{\sqrt{2}}, 0,0\right]$ and $\mathbf{q}_2 = \frac{1}{\sqrt{2}} + \left[0,0,\frac{1}{\sqrt{2}}\right]$, respectively, and their product is simply $\mathbf{q}_1\mathbf{q}_2 = \frac{1}{2} + \left[\frac{1}{2}, \frac{-1}{2}, \frac{1}{2}\right]$. With Rodrigues vectors. $\mathbf{r}_1 = [1,0,0]$ and $\mathbf{r}_2 = [0,0,1]$, respectively, and their product is simply $\mathbf{r}_1\mathbf{r}_2 = [1,-1,1]$. Both $\mathbf{q}_1$ and $\mathbf{r}_1$ represent a rotation of $120°$ around the axis $[1,-1,1]$. The calculation with Rodrigues vector is here simpler than with quaternions. However, things get worse for them if we compose $\mathcal{R}_1$ twice. With quaternions, $\mathbf{q}_1\mathbf{q}_1 = 0 + [1,0,0]$, but with Rodrigues vectors $\mathbf{r}_1\mathbf{r}_2 = \frac{1}{0}[1,0,0]$. Both correspond to a rotation of 180° around the x-axis, but the part 1/0 in the Rodrigues vector generates an error of infinity in computer programs using floating-point numbers. Such numerical issues are not easy to solve, which explains why Rodrigues vectors are not used in industry despite their slight advantage in term of calculation because of this issue.

From the composition law of quaternion defined previously, identity and inversion quaternions can be introduced. The identity quaternion is $\mathbf{e} = 1 + \mathbf{0}$. The inverse of a rotational quaternion is the rotational quaternion of opposite angle, or equivalently of opposite rotation axis, i.e. if $\mathbf{q} = \mathrm{c} + \mathrm{s}\,\tilde{\mathbf{u}}$, then $\mathbf{q}^{-1} = \mathrm{c} - \mathrm{s}\,\tilde{\mathbf{u}}$. Quaternions are very efficient to be compose rotations, but are not appropriate to rotate vectors. Indeed, the image $\mathbf{v}'$ of a vector $\mathbf{v}$ by a rotational quaternion $\mathbf{q} = \mathrm{c} + \mathrm{s}\,\tilde{\mathbf{u}}$ is given by using a product of quaternions

$$\mathbf{q'}_\mathbf{v} = \mathbf{q}\,\mathbf{q}_\mathbf{v}\,\mathbf{q}^{-1} \tag{9}$$

where $\mathbf{q}_\mathbf{v}$ is the quaternion based on $\mathbf{v}$, $\mathbf{q}_\mathbf{v} = 0 + \mathbf{v}$ and $\mathbf{q'}_\mathbf{v}$ is the quaternion based on $\mathbf{v}'$, $\mathbf{q'}_\mathbf{v} = 0 + \mathbf{v}'$. This operation consists actually in using the vector $\mathbf{v}$ to build a rotation of null angle around $\mathbf{v}$ and apply a change of reference basis on this artificially created linear operation $\mathbf{q}_\mathbf{v}$. To be clearer, we write $\boldsymbol{\mathcal{B}}$ a basis and $\boldsymbol{\mathcal{B}}'$ its image by the rotation $\mathbf{q}$. This rotation can be considered as a coordinate transformation matrix from $\boldsymbol{\mathcal{B}}$ to $\boldsymbol{\mathcal{B}}'$, and we note $\mathbf{q} = [\boldsymbol{\mathcal{B}} \to \boldsymbol{\mathcal{B}}']$. The expression of the rotation of null angle around $\mathbf{v}'$ in the basis $\boldsymbol{\mathcal{B}}'$ is the same quaternion as for the rotation of null angle around $\mathbf{v}$ in the basis $\boldsymbol{\mathcal{B}}$, thus we can write $\mathbf{q'}_{\mathbf{v}/\boldsymbol{\mathcal{B}}'} = \mathbf{q}_{\mathbf{v}/\boldsymbol{\mathcal{B}}}$. By the coordinate transformation of linear application, it comes $\mathbf{q'}_{\mathbf{v}/\boldsymbol{\mathcal{B}}} = [\boldsymbol{\mathcal{B}} \to \boldsymbol{\mathcal{B}}']\,\mathbf{q}_{\mathbf{v}/\boldsymbol{\mathcal{B}}}[\boldsymbol{\mathcal{B}}' \to \boldsymbol{\mathcal{B}}]$, which is the rotation of null angle around $\mathbf{v}'$ written in the basis $\boldsymbol{\mathcal{B}}$. Since all the quaternions are written in the basis $\boldsymbol{\mathcal{B}}$, it comes that $\mathbf{q'}_\mathbf{v} = \mathbf{q}\,\mathbf{q}_\mathbf{v}\,\mathbf{q}^{-1}$. Forcing an object (vector) to be transformed into a quaternion (a rotation of null angle) makes the calculations of the images of vectors not efficient. Actually, a most economical way to determine $\mathbf{v}'$ is simply to project the vector $\mathbf{v}$ along the rotation axis $\tilde{\mathbf{u}}$ and write

$\mathbf{v}' = (\tilde{\mathbf{u}} \cdot \mathbf{v})\,\tilde{\mathbf{u}} + \cos(\alpha)(\mathbf{v} - (\tilde{\mathbf{u}} \cdot \mathbf{v})\,\tilde{\mathbf{u}}) + \sin(\alpha)(\tilde{\mathbf{u}} \times \mathbf{v})$, which leads to the formula

$$\mathbf{v}' = \cos(\alpha)\,\mathbf{v} + \big(1 - \cos(\alpha)\big)\,(\tilde{\mathbf{u}} \cdot \mathbf{v})\,\tilde{\mathbf{u}} + \sin(\alpha)\,(\tilde{\mathbf{u}} \times \mathbf{v}) \tag{10}$$

### 1.2. The use of quaternions in EBSD

Electron Back Scatter Diffraction (EBSD) is a technique to acquire orientation maps of the surface of polished samples (Wilkinson & Hirsch, 1997; Schwarzer, 1997; Adams *et al.*, 1993). In each pixel of the EBSD map, the orientation of the crystal $\mathbf{O}$ is encoded by a triplet of Euler angles that constitute the rotation matrix from $\boldsymbol{\mathcal{B}}_S$ a Cartesian basis attached to the surface of sample to $\boldsymbol{\mathcal{B}}_c^{\boxplus}$ a Cartesian basis attached to the crystal, $\mathbf{O} = [\boldsymbol{\mathcal{B}}_S \to \boldsymbol{\mathcal{B}}_c^{\boxplus}]$. The way the Cartesian basis $\boldsymbol{\mathcal{B}}_c^{\boxplus}$ is attached to the crystal basis $\boldsymbol{\mathcal{B}}_c$ is a matter of convention. The associated coordinate transformation matrix $\mathbf{S} = [\boldsymbol{\mathcal{B}}_c^{\boxplus} \to \boldsymbol{\mathcal{B}}_c]$ is called structure tensor in crystallography. It is sometimes noted $\mathbf{A}$ in EBSD (Britton *et al.*, 2016). In most of the EBSD software, the Euler angles of the rotation $\mathbf{O}$ are transformed into quaternions $\mathbf{q}$ to accelerate the calculations. The misorientation between the Cartesian bases attached to the crystals in pixels $i$ and $j$ is $[\boldsymbol{\mathcal{B}}_{ci}^{\boxplus} \to \boldsymbol{\mathcal{B}}_{cj}^{\boxplus}] = [\boldsymbol{\mathcal{B}}_{ci}^{\boxplus} \to \boldsymbol{\mathcal{B}}_S][\boldsymbol{\mathcal{B}}_S \to \boldsymbol{\mathcal{B}}_{cj}^{\boxplus}] = \mathbf{O}_i^{-1}\mathbf{O}_j$ is calculated from the quaternion product $\mathbf{q}_i^{-1}\mathbf{q}_j$. If only the grain boundaries or sub-boundaries have to be plotted in the EBSD, there is no need to determine the rotation axis, and only the scalar part of $\mathbf{q}_i^{-1}\mathbf{q}_j$. The calculations with quaternions are very efficient. However, in a many crystallographic studies, to identify twins in the EBSD maps for example, determining the rotation axes is important, but quaternions become less efficient than matrices. The coordinate of the rotation axis of the misorientation between the pixels $i$ and $j$ is given in Cartesian basis $\boldsymbol{\mathcal{B}}_{ci}^{\boxplus}$ by the vector part of the quaternion $\mathbf{q}_i^{-1}\mathbf{q}_j$, and this vector should be multiplied by the inverse structure tensor to be written in $\boldsymbol{\mathcal{B}}_c$. Using matrix multiplication is indeed more effective than using the quaternion formula (9).

The use of Euler angles and Cartesian quaternions in EBSD relies on a common belief that "*lattice and reciprocal lattice vectors in Cartesian coordinates is required for any kind of metric investigations like calculation of distances, angles, or representation in pole figures*" (Nolze, 2015). It is true that such calculations can be done with the help of the structure tensor, but Cartesian bases are neither required nor effective in term of operations, as it will be explained in next sections. The aim of the paper is to introduce the idea of crystallographic quaternions and propose a generalization of their product law in order to continue to use them directly in the crystal bases without coming back to any Cartesian basis for calculations. Before presenting this law, a new type of crystallographic tensor cousin of metric tensor should be introduced, the cross tensor.

## 2. Metric tensor and Cross tensor

### 2.1. Metric tensor

As mentioned in section 1.2, EBSD uses the structure tensor to go back-and-forth between the crystal basis to a Cartesian basis attached to the crystal. This approach results from the belief that Cartesian bases are required for the calculations of distances or angles between directions or planes, or that the calculations in such bases are more efficient than trying to work directly in the crystal basis. However, it is well-known in crystallography that Cartesian bases and structure tensor are not required at all. The metric tensor is more fundamental than the structure tensor since it does not depend on any arbitrary convention. Let us recall here how the metric tensor is calculated and how to use it. The direct crystal basis $\boldsymbol{\mathcal{B}_c}$ is defined by three non-colinear vectors, $\mathbf{a}, \mathbf{b}, \mathbf{c}$, and we note it $\boldsymbol{\mathcal{B}_c}$. A vector of the direct lattice is defined by its coordinate in $\boldsymbol{\mathcal{B}_c}$ , by $\mathbf{u} = u_a\ \mathbf{a} + u_b\mathbf{b} + u_c\mathbf{c}$. In principle, we should distinguish the vector $\mathbf{u}$ from its set of coordinates in the crystal basis noted $\mathbf{u}_{/\boldsymbol{\mathcal{B}_c}} = \begin{bmatrix} u_a \\ u_b \\ u_c \end{bmatrix}$, but for sake of readability we will often write $\mathbf{u} = \begin{bmatrix} u_a \\ u_b \\ u_c \end{bmatrix}$ for vector of the crystal written in the crystal basis.

The reciprocal basis is $\boldsymbol{\mathcal{B}}_c^*$ ; it is made by the vectors $\mathbf{a}^*,\ \mathbf{b}^*,\ \mathbf{c}^*$, that are such that $\mathbf{a}^* \cdot\ \mathbf{a} = \mathbf{b}^* \cdot\ \mathbf{b} = \mathbf{c}^* \cdot\ \mathbf{c} = 1$, and $\mathbf{a}^* \cdot\ \mathbf{b} = \mathbf{a}^* \cdot\ \mathbf{c} = \mathbf{b}^* \cdot\ \mathbf{c} = 0$, which can be written in a compact matrix form as $(\mathbf{a}^*,\ \mathbf{b}^*,\ \mathbf{c}^*)^{\mathrm{t}}\ (\mathbf{a}, \mathbf{b}, \mathbf{c}) = \mathbf{I}$, where $\mathbf{I}$ is the identity matrix, and "t" means transpose. Let call $\boldsymbol{\mathcal{M}}$ the coordinate transformation matrix $\boldsymbol{\mathcal{M}} = [\boldsymbol{\mathcal{B}}_c^* \to \boldsymbol{\mathcal{B}}_c]$ defined by writing the coordinates of the vectors $\mathbf{a}, \mathbf{b}, \mathbf{c}$ in the basis $\boldsymbol{\mathcal{B}}_c^*$ in columns. Let us find the components of $\boldsymbol{\mathcal{M}}$ by using its properties. We consider two column vectors defined in the direct crystal basis $\boldsymbol{\mathcal{B}_c}$ , by $\mathbf{u} = u_a\ \mathbf{a} + u_b\mathbf{b} + u_c\mathbf{c}$ and $\mathbf{v} = v_a\ \mathbf{a} + v_b\mathbf{b} + v_c\mathbf{c}$ . There are two ways to calculate the scalar product $\mathbf{u} \cdot \mathbf{v}$.

The first way consists in writing the coordinates of the vector $\mathbf{v}$ in the reciprocal basis by using $\boldsymbol{\mathcal{M}}$, by $\mathbf{v}^* = \boldsymbol{\mathcal{M}}\mathbf{v} = v_a^*\ \mathbf{a}^* + v_b^*\ \mathbf{b}^* + v_b^*\ \mathbf{c}^*$. It is then easy to prove that $\mathbf{u} \cdot \mathbf{v} = u_a\ v_a^* + u_b v_b^* + u_c v_c^*$ which is the classical scalar product. This product can be written with different symbols $\mathbf{u} \cdot \mathbf{v} = \mathbf{u} \underline{\cdot} \mathbf{v}^* = \mathbf{u}^{\mathrm{t}}\ \mathbf{v}^* = \mathbf{u}^{\mathrm{t}}\ \boldsymbol{\mathcal{M}}\ \mathbf{v}$. Note that in this paper, the general scalar product is noted with the mark $\cdot$ (whatever the basis, Cartesian or not) and by the mark $\underline{\cdot}$ is used to specify that its is a classical product (sum of the products of the coordinates), or equivalently without any mark for usual matrix-type product of a row vector by a column vector. Note that if $\mathbf{u}$ and $\mathbf{v}$ are defined in a Cartesian basis, $\mathbf{u}^{\mathrm{t}}\ \mathbf{v} = \mathbf{u} \cdot \mathbf{v}$, i.e. a scalar, and $\mathbf{u}\ \mathbf{v}^{\mathrm{t}} = \mathbf{u} \otimes \mathbf{v}$, a matrix obtained by tensorial product.

The second way consists in writing $\mathbf{u} \cdot \mathbf{v} = [u_a, u_b, u_c] \begin{bmatrix} \mathbf{a} \\ \mathbf{b} \\ \mathbf{c} \end{bmatrix} \cdot [\mathbf{a}, \mathbf{b}, \mathbf{c}] \begin{bmatrix} v_a \\ v_b \\ v_c \end{bmatrix} = \mathbf{u}^{\mathrm{t}}\ \boldsymbol{\mathcal{M}}\ \mathbf{v}$. By comparing with the first way, it comes that $\boldsymbol{\mathcal{M}} = \begin{bmatrix} \mathbf{a} \\ \mathbf{b} \\ \mathbf{c} \end{bmatrix} \cdot [\mathbf{a}, \mathbf{b}, \mathbf{c}]$. The matrix $\boldsymbol{\mathcal{M}}$ is formed by the scalar product between the vectors of the crystal basis; it is explicitly

$$\boldsymbol{\mathcal{M}} = \begin{pmatrix} \mathbf{a}^2 & \mathbf{b}^{\mathrm{t}}\,\mathbf{a} & \mathbf{c}^{\mathrm{t}}\,\mathbf{a} \\ \mathbf{a}^{\mathrm{t}}\,\mathbf{b} & \mathbf{b}^2 & \mathbf{c}^{\mathrm{t}}\,\mathbf{b} \\ \mathbf{a}^{\mathrm{t}}\,\mathbf{c} & \mathbf{b}^{\mathrm{t}}\,\mathbf{c} & \mathbf{c}^2 \end{pmatrix} = \begin{pmatrix} \|\mathbf{a}\|^2 & \|\mathbf{b}\|\|\mathbf{a}\|\cos(\gamma) & \|\mathbf{c}\|\|\mathbf{a}\|\cos(\beta) \\ \|\mathbf{a}\|\|\mathbf{b}\|\cos(\gamma) & \|\mathbf{b}\|^2 & \|\mathbf{c}\|\|\mathbf{b}\|\cos(\alpha) \\ \|\mathbf{a}\|\|\mathbf{c}\|\cos(\beta) & \|\mathbf{b}\|\|\mathbf{c}\|\cos(\alpha) & \|\mathbf{c}\|^2 \end{pmatrix} \quad (11)$$

As mentioned previously, the metric tensor allows the calculation of the scalar products between vectors $\mathbf{u}$ and $\mathbf{v}$ by

$$\mathbf{u} \cdot \mathbf{v} = \mathbf{u}^{\mathrm{t}}\ \boldsymbol{\mathcal{M}}\ \mathbf{v} \quad (12)$$

The metric tensor is a coordinate transformation matrix from the reciprocal basis to the direct basis, $\boldsymbol{\mathcal{M}} = [\boldsymbol{\mathcal{B}}_c^* \to \boldsymbol{\mathcal{B}}_c]$ . Its values are given directly from the values of the lattice parameters by equation (11). Fundamentally its calculation requires a measure of length (a ruler) and of angles (a compass). It does not require the use of any intermediate basis. There is no need to write the vectors $\mathbf{a}, \mathbf{b}, \mathbf{c}$ in a Cartesian basis to calculate the metric tensor. It is always possible to use the structure tensor $\mathbf{S}$ to write the vectors **u** and **v** in the same Cartesian basis, then apply classical scalar product, but this way implies unnecessary matrix operations since $\mathbf{u} \cdot \mathbf{v} = (\mathbf{S}\,\mathbf{u}) \underline{\cdot} (\mathbf{S}\,\mathbf{v}) = (\mathbf{S}\,\mathbf{u})^{\mathrm{t}} (\mathbf{S}\,\mathbf{v}) = \mathbf{u}^{\mathrm{t}}\,\mathbf{S}^{\mathrm{t}}\,\mathbf{S}\,\mathbf{v} = \mathbf{u}^{\mathrm{t}}\,\boldsymbol{\mathcal{M}}\,\mathbf{v}$. It is clear that a scalar product is more effective with one matrix-vector product with the metric tensor $\boldsymbol{\mathcal{M}} = \mathbf{S}^{\mathrm{t}}\,\mathbf{S}$, instead of using twice a matrix-vector product with **S**.

The metric tensor is a kind of "scalar-product tensor" since it allows quick calculations of scalar products, and thus of norms and absolute values of angles. What could be its cousin to calculate cross products, what could be a "cross-product tensor"? Surprisingly, we found no trace of such a tensor in the literature of materials science, so let us introduce it here.

### 2.2. Cross tensor

Similarly as for metric tensor, we define the cross tensor $\boldsymbol{\mathcal{X}}$ from its properties. We consider two column vectors defined in the direct crystal basis $\boldsymbol{\mathcal{B}_c}$, by $\mathbf{u} = u_a\,\mathbf{a} + u_b\mathbf{b} + u_c\mathbf{c}$ and $\mathbf{v} = v_a\,\mathbf{a} + v_b\mathbf{b} + v_c\mathbf{c}$. Using the properties that $\mathbf{a} \times \mathbf{a} = \mathbf{0}$, $\mathbf{b} \times \mathbf{b} = \mathbf{0}$ and $\mathbf{c} \times \mathbf{c} = \mathbf{0}$, and that $\mathbf{a} \times \mathbf{b} = -\mathbf{b} \times \mathbf{a}$, $\mathbf{a} \times \mathbf{c} = -\mathbf{c} \times \mathbf{a}$, and $\mathbf{b} \times \mathbf{c} = -\mathbf{c} \times \mathbf{b}$, it comes that the cross product is

$$\mathbf{u} \times \mathbf{v} = (u_a\,\mathbf{a} + u_b\mathbf{b} + u_c\mathbf{c}) \times (v_a\,\mathbf{a} + v_b\mathbf{b} + v_c\mathbf{c})$$

$$= (u_b v_c - u_c v_b)(\mathbf{b} \times \mathbf{c}) - (u_a v_c - u_c v_a)(\mathbf{c} \times \mathbf{a}) + (u_a v_b - u_b v_a)(\mathbf{a} \times \mathbf{b}) =$$

$\boldsymbol{\mathcal{X}} \begin{bmatrix} (u_b v_c - u_c v_b) \\ -(u_a v_c - u_c v_a) \\ (u_a v_b - u_b v_a) \end{bmatrix}$, with the cross tensor $\boldsymbol{\mathcal{X}} = [\mathbf{b} \times \mathbf{c},\ \mathbf{c} \times \mathbf{a},\ \mathbf{a} \times \mathbf{b}]$, a 3x3 matrix obtained by writing in $\boldsymbol{\mathcal{B}_c}$ the coordinates of the three cross product vectors in column. It can also be noted $\boldsymbol{\mathcal{X}} = [\boldsymbol{\mathcal{B}}_c \to \boldsymbol{\mathcal{B}}_c^{\mathcal{X}}]$, with $\boldsymbol{\mathcal{B}}_c^{\mathcal{X}}$ the cross basis constituted by the vectors $\mathbf{b} \times \mathbf{c}$, $\mathbf{c} \times \mathbf{a}$, $\mathbf{a} \times \mathbf{b}$. When the cross tensor is known, the cross product of two vectors is

$$\mathbf{u} \times \mathbf{v} = \boldsymbol{\mathcal{X}}\left(\mathbf{u} \underline{\times} \mathbf{v}\right) \tag{13}$$

where the symbol $\underline{\times}$ means the usual Cartesian cross product obtained by forming a vector with the cross product of the coordinates of the two vectors.

Let us show now that the cross tensor can be obtained from the metric tensor. The three vectors of the reciprocal basis can be obtained from the cross products by $\mathbf{a}^* = \frac{1}{V}(\mathbf{b} \times \mathbf{c})$, $\mathbf{b}^* = \frac{1}{V}(\mathbf{c} \times \mathbf{a})$, $\mathbf{c}^* = \frac{1}{V}(\mathbf{a} \times \mathbf{b})$, with $V$ the volume of the unit cell formed by the three vectors $\mathbf{a}, \mathbf{b}, \mathbf{c}$. The vectors $\mathbf{a}^*, \mathbf{b}^*, \mathbf{c}^*$ constitute the reciprocal basis $\boldsymbol{\mathcal{B}}_c^* = (\mathbf{a}^*, \mathbf{b}^*, \mathbf{c}^*)$. The cross and reciprocal bases are thus linked by

$$\boldsymbol{\mathcal{B}}_c^{\mathcal{X}} = V\,\boldsymbol{\mathcal{B}}_c^* \tag{14}$$

Note that the volume of the unit cell is $V = \det(\boldsymbol{a}, \boldsymbol{b}, \boldsymbol{c})$, the determinant of the matrix formed by the crystallographic vectors written in a Cartesian basis, i.e. $V = \det(\boldsymbol{S})$ with $\boldsymbol{S}$ the structure tensor, but it can also be deduced directly from the metric tensor. Indeed, since $\boldsymbol{\mathcal{M}} = \mathbf{S}^{\mathrm{t}}\,\mathbf{S}$, $\det(\boldsymbol{\mathcal{M}}) = V^2$, thus

$$V = \sqrt{\det(\boldsymbol{\mathcal{M}})} \tag{15}$$

Using $\boldsymbol{\mathcal{M}} = [\boldsymbol{\mathcal{B}}_c^* \to \boldsymbol{\mathcal{B}}_c]$, $\boldsymbol{\mathcal{M}}^{-1} = [\boldsymbol{\mathcal{B}}_c \to \boldsymbol{\mathcal{B}}_c^*]$, and $\boldsymbol{\mathcal{X}} = [\boldsymbol{\mathcal{B}}_c \to \boldsymbol{\mathcal{B}}_c^{\mathcal{X}}]$ it comes immediately that $\boldsymbol{\mathcal{X}} = V\,\boldsymbol{\mathcal{M}}^{-1}$. The cross tensor can thus be written from the metric tensor only by:

$$\boldsymbol{\mathcal{X}} = \sqrt{\det(\boldsymbol{\mathcal{M}})}\,\boldsymbol{\mathcal{M}}^{-1} \qquad (16)$$

## 3. Crystallographic quaternions

It is natural to generalize the rotational quaternions defined for Cartesian bases by the same objects that now refer to the crystal basis $\boldsymbol{\mathcal{B}}_c$. For a rotation of angle α around the axis $\mathbf{u}$ written in $\boldsymbol{\mathcal{B}}_c$ the associated crystallographic quaternion $\mathbf{q}$ is simply defined by

$$\mathbf{q} = \mathrm{c} + \mathrm{s}\,\tilde{\mathbf{u}} \qquad (17)$$

with c and s the cosine and sine of the semi-angle of the rotation, $\mathrm{c} = \cos(\frac{\alpha}{2})$ and $\mathrm{s} = \sin(\frac{\alpha}{2})$, and the tilde symbol over $\mathbf{u}$ means that this vector is normalized in $\boldsymbol{\mathcal{B}}_c$, i.e. $\|\tilde{\mathbf{u}}\| = \sqrt{\tilde{\mathbf{u}}^t\,\boldsymbol{\mathcal{M}}\,\tilde{\mathbf{u}}} = 1$.

If two rotations $\mathcal{R}_1$ and $\mathcal{R}_2$ are represented by the crystallographic quaternions $\mathbf{q}_1 = \mathrm{c}_1 + \mathrm{s}_1\,\widetilde{\mathbf{u}_1}$ and $\mathbf{q}_2 = \mathrm{c}_2 + \mathrm{s}_2\,\widetilde{\mathbf{u}_2}$ , respectively, their composition is given by the crystallographic quaternion $\mathbf{q}_3$ obtained by generalizing the scalar product for the real component $\mathrm{c}_3 = \mathrm{c}_1\mathrm{c}_2 - \mathrm{s}_1\mathrm{s}_2\,\widetilde{\mathbf{u}_1}.\widetilde{\mathbf{u}_2}$, and the cross product for the vectorial component $\mathrm{s}_3\,\widetilde{\mathbf{u}_3} = \mathrm{c}_2\,\mathrm{s}_1\,\widetilde{\mathbf{u}_1} + \mathrm{c}_1\mathrm{s}_2\,\widetilde{\mathbf{u}_2} + \mathrm{s}_1\mathrm{s}_2\,\widetilde{\mathbf{u}_1} \times \widetilde{\mathbf{u}_2}$ of equations (3) and (4), respectively. From equations (12) and (13), it comes that

$$\mathbf{q}_3 = \mathbf{q}_1\mathbf{q}_2 = \mathrm{c}_3 + \mathrm{s}_3\,\widetilde{\mathbf{u}_3} \quad \text{with} \qquad (18)$$

$$\mathrm{c}_3 = \mathrm{c}_1\mathrm{c}_2 - \mathrm{s}_1\mathrm{s}_2\left(\widetilde{\mathbf{u}_1}^{t}\boldsymbol{\mathcal{M}}\,\widetilde{\mathbf{u}_2}\right) \qquad (19)$$

$$\mathrm{s}_3\,\widetilde{\mathbf{u}_3} = \mathrm{c}_2\,\mathrm{s}_1\,\widetilde{\mathbf{u}_1} + \mathrm{c}_1\mathrm{s}_2\,\widetilde{\mathbf{u}_2} + \mathrm{s}_1\mathrm{s}_2\,\boldsymbol{\mathcal{X}}\left(\widetilde{\mathbf{u}_1} \underline{\times} \widetilde{\mathbf{u}_2}\right) \qquad (20)$$

with the cross tensor $\boldsymbol{\mathcal{X}}$ directly deduced from the metric tensor $\boldsymbol{\mathcal{M}}$ by equation (16), and $\underline{\times}$ the usual Cartesian cross product.

In their compact form, the composition of two quaternions $\mathbf{q}_1 = \mathrm{c}_1 + \mathbf{u}_1$ by $\mathbf{q}_2 = \mathrm{c}_2 + \mathbf{u}_2$ is

$$\mathbf{q}_3 = \mathbf{q}_1\mathbf{q}_2 = \mathrm{c}_3 + \mathbf{u}_3 \quad \text{with} \qquad (21)$$

$$\mathrm{c}_3 = \mathrm{c}_1\mathrm{c}_2 - (\mathbf{u}_1^t\,\boldsymbol{\mathcal{M}}\,\mathbf{u}_2) \qquad (22)$$

$$\mathbf{u}_3 = \mathrm{c}_2\,\mathbf{u}_1 + \mathrm{c}_1\mathbf{u}_2 + \boldsymbol{\mathcal{X}}\left(\mathbf{u}_1 \underline{\times}\, \mathbf{u}_2\right) \qquad (23)$$

The composition of Rodrigues vectors $\mathbf{r}_1$ and $\mathbf{r}_2$ given in the crystal basis is also generalized by

$$\mathbf{r}_3 = \mathbf{r}_1\,\mathbf{r}_2 = \left(\frac{1}{1-\mathbf{r}_1^t\boldsymbol{\mathcal{M}}\,\mathbf{r}_2}\right)(\mathbf{r}_1 + \mathbf{r}_2 + \boldsymbol{\mathcal{X}}\left(\mathbf{r}_1 \underline{\times}\, \mathbf{r}_2\right)) \qquad (24)$$

A computer program that calculates the different types of products (Cartesian quaternions, crystallographic quaternions, and Rodrigues) for any crystal system has been written. It is available upon request to the author.

The metric and cross tensors can also be used to calculate the image $\mathbf{v}'$ of a vector $\mathbf{v}$ by a rotation of angle α around a rotation axis $\tilde{\mathbf{u}}$ by a formula that generalized equation (10), i.e.

$$\mathbf{v}' = \cos(\alpha)\,\mathbf{v} + \left(1 - \cos(\alpha)\right)(\tilde{\mathbf{u}}^t\,\boldsymbol{\mathcal{M}}\,\mathbf{v})\,\tilde{\mathbf{u}} + \sin(\alpha)\,\boldsymbol{\mathcal{X}}\left(\tilde{\mathbf{u}} \underline{\times}\, \mathbf{v}\right) \qquad (25)$$

## 4. Possible application of crystallographic quaternions to EBSD

As detailed in section 1.2, Cartesian rotational quaternions are used in EBSD to encode the orientations of the grains by triplets of Euler angles, even if their phase is not cubic. The "trick" consists in using the structure tensor as coordinate transformation matrix from a Cartesian basis attached to the crystal to the crystal basis. The Cartesian basis is defined from arbitrary conventions. The structure tensor is not fundamental, contrarily to the metric tensor. We have seen in the previous section that there is no need to introduce any Cartesian basis to define and manipulate rotational quaternions for a crystal directly in its crystal basis. Crystallographic quaternions can be defined exactly as usual Cartesian quaternions, and only their composition law has to adapted by generalizing the scalar product and the cross product to non-Cartesian bases. The use of the metric tensor to calculate scalar product was well known, but to our knowledge, the introduction of the cross tensor is new. Now, a practical question comes: is the calculation of the composition of two rotations of non-cubic crystals more efficient by using the crystallographic quaternions allow or by using Cartesian quaternions and back-and forth transformations between the Cartesian and crystal bases? This question is important because if the answer is positive, it may change the way we treat the EBSD maps.

EBSD maps are obtained by acquiring on a camera the Kikuchi diffraction patterns of the backscatter electrons. The Kikuchi bands are classically identified by Hough transforms. After corrections by some projection factors of the camera, the angles between the bands are determined and compared with the angles between the crystallographic planes $(hkl)$. When a good match is found, the bands are identified to $(hkl)$ planes, and the orientation of the crystal relatively to the sample surface is determined. More recently, the experimental Kikuchi patterns are directly compared to pre-calculated 2D master Kikuchi patterns by a method called "pattern matching" (Winkelmann *et al.*, 2020). Even more effective and promising is the "spherical indexing" method that works by projecting the experimental Kikuchi patterns onto a sphere, and by cross-correlating them with a unique pre-calculated 3D spherical master pattern of the crystal (Day, 2008; Lenthe *et al.*, 2020). The rotation axis and rotation angle encoding the orientation in each pixel of the map is given relatively to the spherical master pattern, i.e. directly in the crystal basis. All the orientations can thus be directly encoded by crystallographic quaternions written in the crystal basis of the master pattern. The misorientations can then be calculated directly from these quaternions. Only final operations, such as plotting pole figures of crystallographic directions or planes in the Cartesian reference frame of the surface or of the sample, or plotting the inverse pole figures of some specific directions of the sample into the crystal basis, requires to know the coordinate transformation matrix between the sample or the surface and the crystal basis of the spherical master pattern. Since the experimental EBSD data are intrinsically given by crystallographic quaternions, it seems more effective to continue to use them instead of going back-and-forth to Cartesian bases. Let us show it. In the following, we will use FLOP for "floating-point operation", and the letters M for "multiplication" and A for "addition".

For cubic crystals, the product of crystallographic quaternions is identical to that given by the product of Cartesian quaternions. Indeed, $\boldsymbol{\mathcal{M}} = a^2\,\mathbf{I}$ with $a$ the lattice parameter of the cubic phase, and the two rotation axes $\tilde{\mathbf{u}}$ that compose the quaternions have a crystallographic norm $\sqrt{\tilde{\mathbf{u}} \cdot \tilde{\mathbf{u}}} = \sqrt{\tilde{\mathbf{u}}^{\mathrm{t}}\, \boldsymbol{\mathcal{M}}\, \tilde{\mathbf{u}}} = a\sqrt{\tilde{\mathbf{u}} \underset{-}{\cdot} \tilde{\mathbf{u}}} = 1$, that is thus $\frac{1}{a}$ times the Cartesian norm. To avoid, unnecessary multiplication and division by $a$, the unit can be changed such that $a = 1$ (in arbitrary unit). Then, $\boldsymbol{\mathcal{M}} = \mathbf{I}$ and the crystallographic product is identical to the Cartesian product. By writing the rotational Cartesian quaternions compactly with $\mathbf{q} = \mathrm{c} + \mathbf{u}$, one can check that the calculation of the angular part of the product of two quaternions in equation (6) requires 4 M and 4 A, and that of the vectorial part in equation (7) requires 12 M and 6 A. Thus, the product of two rotational Cartesian quaternions requires a total of 28 FLOPs (16 M and 12 A).

For non-cubic crystals, the number of operations required to calculate the product of two crystallographic quaternions can be determined from formula (18). It is important to note that the product of a symmetric matrix by a vector requires only 9 FLOPS (6 M and 3 A) with optimized algorithms (Stewart, 1973), and not 15 FLOPS (9 M and 6 A) as for usual matrices. Consequently, the calculation of the angular and vectorial parts of the product requires 12 M and 6 A in addition to that of the usual Cartesian quaternion product, which involves a total of 46 FLOPs (28 M and 18 A). This number of operations is higher than with Cartesian quaternions (28 FLOPs).

From this first rough comparison, one could conclude that the crystallographic quaternions would make the treatment of the EBSD slower in comparison with the usual methods based on the use of the Cartesian quaternions deduced from by the Euler angles. However, for a fair comparison, it is important to take into account that additional operations are required in these methods:

(a) The coordinates of the rotation axis in the Cartesian basis have to be transformed into those of the crystal basis, which requires the use of the structure tensor, and thus additional 15 FLOPs (9 M and 6 A). Therefore, using Cartesian quaternions with a change of coordinate of the rotation axis requires 43 FLOPs. This number is still lower than with crystallographic quaternions (total 46 FLOPs), but other points should be considered.

(b) We assumed that the quaternions were already determined and encoded in the EBSD map, which does not take into account the step made during the acquisition of the map during which the crystallographic data extracted from the Kikuchi patterns, such as the axis $[uvw]$ of the pattern centre and a reference plane $(hkl)$, are converted into triplets of Euler angles. Such calculations imply to multiply the orientation matrix by the structure tensor (or its inverse), i.e. additional 45 FLOPs (9 M and 6 A). Thus, if the full process is considered, from the acquisition of the Kikuchi patterns to the writing the quaternions in a EBSD file for each pixel, using crystallographic quaternions seems to be more effective than using Cartesian quaternions. With crystallographic quaternions, all the calculations are performed in the crystal basis, which avoids going back-and forth between the crystal basis and an arbitrary Cartesian basis attached to the crystal.

(c) The symmetry operations take simple form in terms of crystallographic quaternions because the symmetry axes are always simply written in the crystal basis (their coordinates are 0, ±1) and the cosines are always an integer between -1 and 2, and the sines always $\pm\frac{\sqrt{3}}{2}$, 0, or ±1.

The expressions of symmetries with crystallographic quaternions is thus simpler and more elegant than with Cartesian quaternions. Recognizing a possible symmetry operation between two pixels, which means an identical orientation of these pixels in an EBSD is thus more effective with crystallographic quaternions than with Cartesian quaternions.

(d) The structure tensor and the metric tensor contain some null values and unit values (if the lattice parameter a is taken as the unit reference) for highly symmetric crystals, which reduces the number of FLOPs, but the effects of this reduction is difficult to compare. For example, for a hexagonal crystal of lattice parameters $a = 1$ and $c$, the conventional structure tensor is $\mathbf{S} = \begin{bmatrix} 1 & \frac{-1}{2} & 0 \\ 0 & \frac{\sqrt{3}}{2} & 0 \\ 0 & 0 & c \end{bmatrix}$ and the metric tensor is $\boldsymbol{\mathcal{M}} = \begin{bmatrix} 1 & \frac{-1}{2} & 0 \\ \frac{-1}{2} & 1 & 0 \\ 0 & 0 & c \end{bmatrix}$. Both contain 6 values equal to 0 or 1. We note that the metric tensor is symmetric and contain only rational values (besides $c$) whereas the structure tensor contains the irrational value $\sqrt{3}$ , but since the calculations made with the commercial EBSD program uses floating-point numbers, no difference of speed is expected between calculations with $\mathbf{S}$ or with $\boldsymbol{\mathcal{M}}$. However, it should be noted that the convention on $\mathbf{S}$ depends on the EBSD company (Bruker and Oxford Instruments do not use the same convention as EDAX), whereas the metric tensor is independent of any convention.

## 5. Conclusions

Unit quaternions are widely used to encode rotations. However, in their usual form, quaternions refer to a Cartesian basis, so they cannot be used directly in crystallography to determine the rotations and their products in non-cubic crystals. In general, back-and-forth changes between the crystal basis and a Cartesian basis attached to the crystal are applied. This study shows that actually such changes of bases are not required and that quaternions can be used directly in the crystal basis if the quaternion product formula is generalized. Since the usual formula contains a scalar product and a cross product, these two types of products were adapted to the crystal basis. The metric tensor responds to this task for the scalar product, but, to the author knowledge, there were no equivalent for the cross product. Thus, a matrix called "cross tensor" was introduced; it transforms the usual Cartesian cross product into a crystallographic cross product. It was then showed that the cross tensor is proportional to the inverse of the metric tensor by equation (16). The formula for the product of two crystallographic quaternion product was then directly deduced; it depends uniquely on the metric tensor by equations (18)-(20). We think that the crystallographic quaternions could advantageously replace the Cartesian quaternions in the computer programs of Electron Back Scatter Diffraction.

**Table 1** Comparisons of the number of FLOPs (floating-point operations) for different types of products in a 3D space. In this table, M means "multiplication", and A "addition" or subtraction.

| | Number of M | Number of A | Number of FLOPs |
|---|---|---|---|
| Cartesian scalar product of two vectors | 3 | 2 | 5 |
| Cartesian cross product of two vectors | 6 | 3 | 9 |
| **Cartesian quaternion product** | **16** | **12** | **28** |
| **Rodrigues product** | **12** | **12** | **24** |
| Matrix - vector product | 9 | 6 | 15 |
| Symmetric matrix - vector product | 6 | 3 | 9 |
| **Crystallographic quaternion product** | **28** | **18** | **46** |
| **Crystallographic Rodrigues product** | **24** | **18** | **42** |

**Acknowledgements** I acknowledge Prof. Roland Logé, my boss, chief of the lab LMTM, and PX group, the company that chaired the lab for ten years.

**Appendix 1. The composition law of quaternions and the spherical laws**

A rotation is the composition of two mirror planes (tail, head) such that these planes intersect along the rotation axis, and the angle between them is the semiangle of the rotation angle. The composition of two rotations can thus be geometrically constructed by decomposing them such that the head mirror plane of the first rotation is the tail mirror plane of the second rotation, which is a usual groupoid composition law (Cayron, 2006). The composition of two rotations can thus be illustrated as follows. We consider three vectors $\mathbf{u}_1$, $\mathbf{u}_2$ and $\mathbf{u}_3$ of unit length. Their pairs define the planes $\mathbf{p}_3 = (\mathbf{u}_1, \mathbf{u}_2)$, $\mathbf{p}_1 = (\mathbf{u}_2, \mathbf{u}_3)$, and $\mathbf{p}_2 = (\mathbf{u}_3, \mathbf{u}_1)$, as represented in Figure A1(a). The order of the three indices is important and should belong to the group of the cyclic permutations $(1,2,3)$. In order to facilitate the notations, we note $\|\mathbf{u}_i\| = \mathrm{u}_i$ (the vector is in bold and its norm is regular). We also note $u_{ij}$ the angle between the vectors $\mathbf{u}_i$ and $\mathbf{u}_j$, and $p_{ij}$ the angle between the planes $\mathbf{p}_i$ and $\mathbf{p}_j$. We call $\mathcal{R}_1$ the rotation around $\mathbf{u}_1$ of angle $2p_{23}$, $\mathcal{R}_2$ the rotation around $\mathbf{u}_2$ of angle $2p_{31}$, $\mathcal{R}_3$ the rotation around $\mathbf{u}_3$ of angle $2p_{21}$. By construction, $\mathcal{R}_3$ is the composition of $\mathcal{R}_1$ followed by $\mathcal{R}_2$.

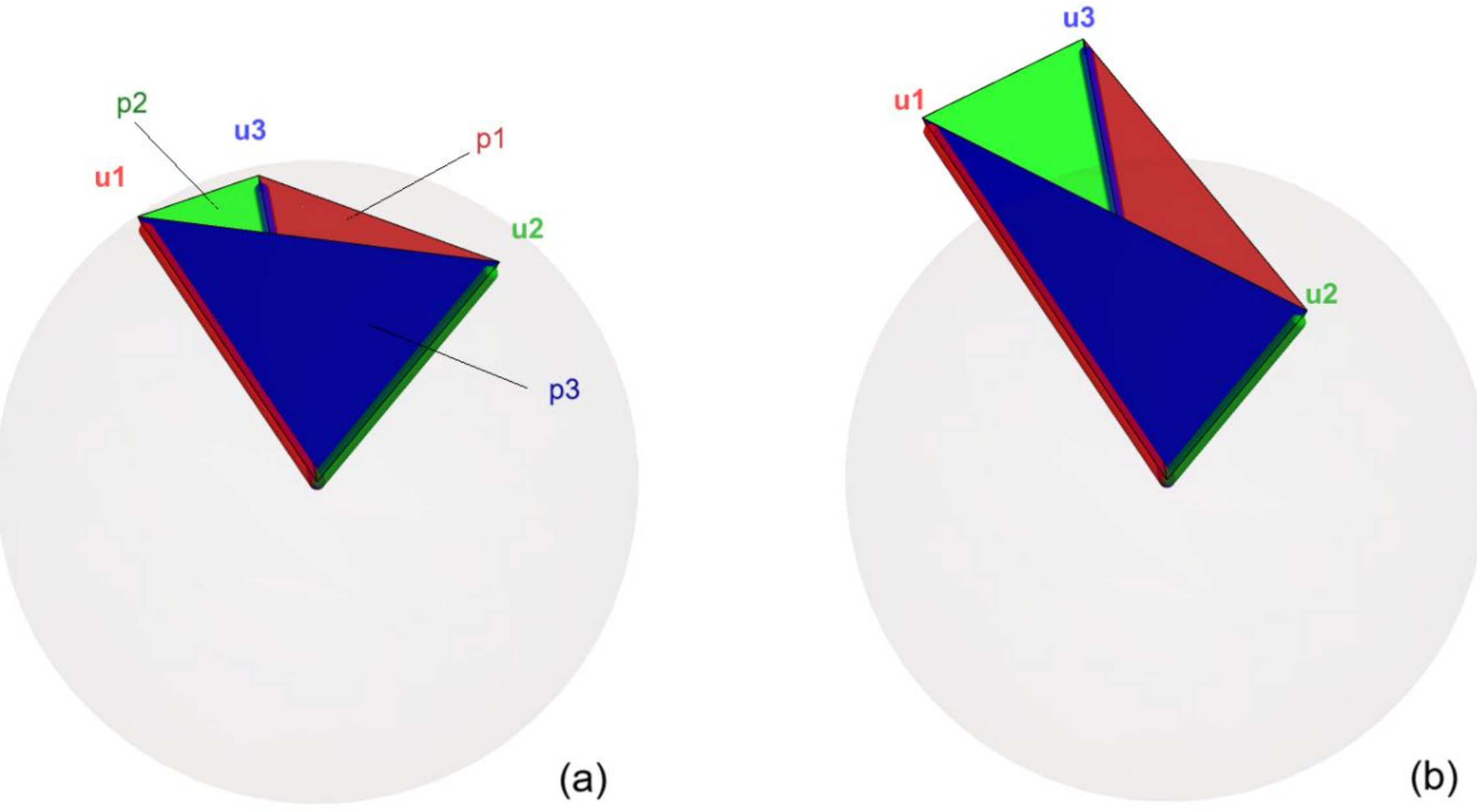

Figure A1. Three vectors and the planes formed by paring the vectors. This configuration is used to represent the quaternion $\mathbf{q}_1$, $\mathbf{q}_2$ and $\mathbf{q}_3$. The quaternion $\mathbf{q}_i$ represents the rotation around the axis $\mathbf{u}_i$ with rotation angle $2p_{jk}$, with $p_{jk}$ the angle between the planes $\mathbf{p}_j$ and $\mathbf{p}_k$. The three indices $\{i, j, k\} \in (1,2,3)$. (a) The rotation vectors $\mathbf{u}_i$ are unit vectors. They point to the unit sphere in grey. (b) Same configuration, with the same directions of the vectors and planes, but now the norms of $\mathbf{u}_i$ have been changed such that cross-product vectors $\mathbf{u}_i \times \mathbf{u}_j$ are unit vectors, i.e. the surfaces of the planes in blue, green and red are equal to 1.

Writing the three rotations $\mathcal{R}_i$ by their quaternions $\mathbf{q}_i = \mathrm{c}_i + \mathrm{s}_i \, \widetilde{\mathbf{u}_i}$ , with $\mathrm{c}_i = \cos(p_{jk})$ and $\mathrm{c}_i = \sin(p_{jk})$. The composition of $\mathbf{q}_1 = \mathrm{c}_1 + \mathrm{s}_1 \, \widetilde{\mathbf{u}_1}$ followed by $\mathbf{q}_2 = \mathrm{c}_2 + \mathrm{s}_2 \, \widetilde{\mathbf{u}_2}$ is $\mathbf{q}_3 = \mathbf{q}_1\mathbf{q}_2 = \mathrm{c}_3 + \mathrm{s}_3 \, \widetilde{\mathbf{u}_3}$ with $\mathrm{c}_3 = \mathrm{c}_1\mathrm{c}_2 - \mathrm{s}_1\mathrm{s}_2 \, \widetilde{\mathbf{u}_1}.\widetilde{\mathbf{u}_2}$ and $\mathrm{s}_3 \, \widetilde{\mathbf{u}_3} = \mathrm{c}_2 \, \mathrm{s}_1 \, \widetilde{\mathbf{u}_1} + \mathrm{c}_1\mathrm{s}_2 \, \widetilde{\mathbf{u}_2} + \mathrm{s}_1\mathrm{s}_2 \, (\widetilde{\mathbf{u}_1} \times \widetilde{\mathbf{u}_2})$.

The formula of angular part $\mathrm{c}_3$ results from the spherical law of cosines for the planes. Indeed, as it will be shown and demonstrated in the next Appendix, this law says that

$$\cos(p_{12}) = \cos(p_{23})\cos(p_{31}) - \sin(p_{23})\sin(p_{31})\cos(u_{12}), \text{ i.e. } \mathrm{c}_3 = \mathrm{c}_1\mathrm{c}_2 - \mathrm{s}_1\mathrm{s}_2 \, \widetilde{\mathbf{u}_1}.\widetilde{\mathbf{u}_2}.$$

The formula of axial part $\mathrm{s}_3 \, \widetilde{\mathbf{u}_3}$ can be demonstrated by the spherical law of sines. We use the same configuration of axes $\mathbf{u}_i$ and planes $\mathbf{p}_i$to define the three rotations, except that now we chose the lengths of the vectors $\mathbf{u}_i$ such that their cross-products are units, i.e. $\|\mathbf{u}_j \times \mathbf{u}_k\| = \mathrm{u}_j\mathrm{u}_k \sin(u_{jk}) = 1$. This is illustrated in Figure A1(b). The lengths of the vectors $\mathbf{u}_i$ are given in the next appendix by equation (A9). We decompose $\mathbf{u}_3$ in the basis formed by the vectors $\mathbf{u}_1, \mathbf{u}_2$ and $\mathbf{u}_1 \times \mathbf{u}_2$ by writing $\mathbf{u}_3 = x \, \mathbf{u}_1 + y \, \mathbf{u}_2 + z \, (\mathbf{u}_1 \times \mathbf{u}_2)$. We use $p_{ij}$ the angle between the planes $\mathbf{p}_i$ and $\mathbf{p}_j$ to determine the coefficients $x, y, z$.

For the planes $\mathbf{p}_1$ and $\mathbf{p}_3$ we get

$$c_2 = \cos(p_{13}) = (\mathbf{u}_2 \times \mathbf{u}_3) \cdot (\mathbf{u}_1 \times \mathbf{u}_2) = \left(-x \, (\mathbf{u}_2 \times \mathbf{u}_1) + z \, \mathbf{u}_2 \times (\mathbf{u}_1 \times \mathbf{u}_2)\right) \cdot (\mathbf{u}_1 \times \mathbf{u}_2) = x$$

For the planes $\mathbf{p}_2$ and $\mathbf{p}_3$ we get

$$c_1 = \cos(p_{23}) = (\mathbf{u}_3 \times \mathbf{u}_1) \cdot (\mathbf{u}_1 \times \mathbf{u}_2) = (-y \, (\mathbf{u}_2 \times \mathbf{u}_1) + z \, (\mathbf{u}_1 \times \mathbf{u}_2) \times \mathbf{u}_1) \cdot (\mathbf{u}_1 \times \mathbf{u}_2) = y$$

To find $z$, we write $\mathbf{u}_3 \cdot (\mathbf{u}_1 \times \mathbf{u}_2) = V = z$.

It comes that $\mathbf{u}_3 = \mathrm{c}_2 \, \mathbf{u}_1 + \mathrm{c}_1\mathbf{u}_2 + V(\mathbf{u}_1 \times \mathbf{u}_2)$

We write it $\mathrm{u}_3\widetilde{\mathbf{u}_3} = \mathrm{c}_2 \, \mathrm{u}_1\widetilde{\mathbf{u}_1} + \mathrm{c}_1 \, \mathrm{u}_2\widetilde{\mathbf{u}_2} + V(\mathrm{u}_1\widetilde{\mathbf{u}_1} \times \mathrm{u}_2\widetilde{\mathbf{u}_2})$

By using equation (A10) it comes directly that $\mathrm{s}_3\widetilde{\mathbf{u}_3} = \mathrm{c}_2 \, \mathrm{s}_1\widetilde{\mathbf{u}_1} + \mathrm{c}_1 \, \mathrm{s}_2\widetilde{\mathbf{u}_2} + \mathrm{s}_1 \, \mathrm{s}_2(\widetilde{\mathbf{u}_1} \times \widetilde{\mathbf{u}_2})$, which is the axial part of the product to the quaternions.

## Appendix 2. The spherical laws of cosines and sines demonstrated by metrics

There are many proofs of the spherical laws of cosines and sines, and all of them are based on the calculations of scalar product and/or cross product of unit vectors. We give here a proof based on similar arguments, but written in a crystallography style. As in the previous appendix we consider three vectors $\mathbf{u}_1$, $\mathbf{u}_2$ and $\mathbf{u}_3$ of same length, and their pairs defining the planes $\mathbf{p}_3 = (\mathbf{u}_1, \mathbf{u}_2)$, $\mathbf{p}_1 = (\mathbf{u}_2, \mathbf{u}_3)$, and $\mathbf{p}_2 = (\mathbf{u}_3, \mathbf{u}_1)$, as represented in Figure A1. We continue to call $u_{ij}$ the angles between the vectors $\mathbf{u}_i$ and $\mathbf{u}_j$, and $p_{ij}$ the angles between the planes $\mathbf{p}_i$ and $\mathbf{p}_j$. The vectors $\mathbf{u}_1$, $\mathbf{u}_2$ and $\mathbf{u}_3$ form the unit cell of a lattice. The components of the metric of this lattice are the cross products $\mathbf{u}_i \cdot \mathbf{u}_j = \|\mathbf{u}_i\|\|\mathbf{u}_j\| \cos(u_{ij})$. As we assumed that the norm of the vectors is 1, i.e. $\|\mathbf{u}_i\| = 1$, the metric tensor is

$$\boldsymbol{\mathcal{M}}_{\boldsymbol{u}} = \begin{pmatrix} 1 & \cos(u_{12}) & \cos(u_{13}) \\ \cos(u_{21}) & 1 & \cos(u_{23}) \\ \cos(u_{31}) & \cos(u_{32}) & 1 \end{pmatrix} \tag{A1}$$

The reciprocal lattice is formed by the planes $\mathbf{p}_1$, $\mathbf{p}_2$ and $\mathbf{p}_3$. The norm of $\mathbf{p}_i$ in this reciprocal space is the inverse of the interplanar distance $h_i$ of the plane $\mathbf{p}_i$ in the lattice; it is

$\|\mathbf{p}_i\| = \frac{1}{h_i} = \frac{Surface(\mathbf{u}_j,\mathbf{u}_k)}{V} = \frac{\|\mathbf{u}_j \times \mathbf{u}_k\|}{V} = \frac{\sin(u_{jk})}{V}$, where $\{i,j,k\} \in (1,2,3)$, with $(1,2,3)$ the cyclic permutations of the three indices. The part $Surface(\mathbf{u}_j, \mathbf{u}_k)$ is the surface of the rhombus formed by the two vectors, and $V = \det(\mathbf{u}_1, \mathbf{u}_2, \mathbf{u}_3)$ is the volume of the unit cell. The components of the metric tensor of the reciprocal lattice are the cross products $\mathbf{p}_i \cdot \mathbf{p}_j = \|\mathbf{p}_i\| \, \|\mathbf{p}_j\| \cos(p_{ij})$

$$\boldsymbol{\mathcal{M}}_u^* = \frac{1}{V^2}\begin{pmatrix} \sin^2(u_{23}) & \sin(u_{13})\sin(u_{32})\cos(p_{12}) & \sin(u_{12})\sin(u_{23})\cos(p_{13}) \\ \sin(u_{23})\sin(u_{31})\cos(p_{21}) & \sin^2(u_{13}) & \sin(u_{21})\sin(u_{13})\cos(p_{23}) \\ \sin(u_{32})\sin(u_{21})\cos(p_{31}) & \sin(u_{31})\sin(u_{12})\cos(p_{32}) & \sin^2(u_{12}) \end{pmatrix} \tag{A2}$$

We used the properties that $\sin(u_{ij})\sin(u_{jk}) = \sin(u_{kj})\sin(u_{ki})$ to write the components in position $(i,j)$ in the form $\sin(u_{ik})\sin(u_{kj})\cos(p_{ij})$. It can be checked that $\boldsymbol{\mathcal{M}}_u^*$ is symmetric since $u_{ij} = -u_{ji}$.

The spherical law of angles directly comes from the equality $\boldsymbol{\mathcal{M}}_u^* = \boldsymbol{\mathcal{M}}_u^{-1}$.

Indeed, $\boldsymbol{\mathcal{M}}_u^{-1} = \frac{\mathrm{adj}(\boldsymbol{\mathcal{M}}_{\boldsymbol{u}})}{\det(\boldsymbol{\mathcal{M}}_{\boldsymbol{u}})}$ with $\det(\boldsymbol{\mathcal{M}}_{\boldsymbol{u}}) = V^2$, and $\mathrm{adj}(\boldsymbol{\mathcal{M}}_{\boldsymbol{u}})$ is the adjugate matrix of $\boldsymbol{\mathcal{M}}_{\boldsymbol{u}}$, i.e. the transpose of cofactor matrix, it comes that

$$\boldsymbol{\mathcal{M}}_u^{-1} = \frac{1}{V^2}\begin{pmatrix} 1-\cos^2(u_{23}) & \cos(u_{13})\cos(u_{32}) - \cos(u_{12}) & \cos(u_{12})\cos(u_{23}) - \cos(u_{13}) \\ \cos(u_{23})\cos(u_{31}) - \cos(u_{21}) & 1-\cos^2(u_{13}) & \cos(u_{12})\cos(u_{13}) - \cos(u_{23}) \\ \cos(u_{32})\cos(u_{21}) - \cos(u_{31}) & \cos(u_{31})\cos(u_{12}) - \cos(u_{32}) & 1-\cos^2(u_{12}) \end{pmatrix} \tag{A3}$$

We used the properties that $\cos(u_{ij}) = \cos(u_{ji})$ to write the components in position $(i,j)$ in the form $\cos(u_{ik})\cos(u_{kj}) - \cos(u_{ij})$ where $\{i,j,k\} \in (1,2,3)$.

The equality of the components of $\boldsymbol{\mathcal{M}}_u^*$ and $\boldsymbol{\mathcal{M}}_u^{-1}$ becomes

$\cos(u_{ik})\cos(u_{kj}) - \cos(u_{ij}) = \sin(u_{ik})\sin(u_{kj})\cos(p_{ij})$, or equivalently

$$\cos(u_{ij}) = \cos(u_{ik})\cos(u_{kj}) - \sin(u_{ik})\sin(u_{kj})\cos(p_{ij}) \tag{A4}$$

Noting the angles between the directions $u_{12} = c$, $u_{23} = a$ and $u_{31} = b$, and the angles between the planes (dihedral angles) $p_{12} = C$, $p_{23} = A$ and $p_{31} = B$, and the equation becomes

$$\cos(c) = \cos(a)\cos(b) - \sin(a)\sin(b)\cos(C) \tag{A5}$$

It is the spherical law of cosines for the directions. Note that this law is often written with a sign $+$ instead of $-$ because the letter $b$ is (oddly) attributed to $u_{13}$ instead of $u_{31}$.

The same method can be used by assuming now that the norm of the reciprocal vectors is 1, i.e. $\|\mathbf{p}_i\| = 1$. The metrics of the planes is

$$\boldsymbol{\mathcal{M}_p} = \begin{pmatrix} 1 & \cos(p_{12}) & \cos(p_{13}) \\ \cos(p_{21}) & 1 & \cos(p_{23}) \\ \cos(p_{31}) & \cos(p_{32}) & 1 \end{pmatrix} \tag{A6}$$

$\boldsymbol{\mathcal{M}_p}$ is symmetric. Applying the same arguments as previously used for the directions, and now using $\boldsymbol{\mathcal{M}_p}$ and $\boldsymbol{\mathcal{M}}_p^*$, we obtain the law of cosines for the planes,

$\cos(p_{ik})\cos(p_{kj}) - \cos(p_{ij}) = \sin(p_{ik})\sin(p_{kj})\cos(u_{ij})$, or equivalently

$$\cos(p_{ij}) = \cos(p_{ik})\cos(p_{kj}) - \sin(p_{ik})\sin(p_{kj})\cos(u_{ij}) \tag{A7}$$

With the notations as used for equation (A5), it becomes

$$\cos(C) = \cos(A)\cos(B) - \sin(A)\sin(B)\cos(c) \tag{A8}$$

It is the spherical law of cosines for the planes.

Now, we demonstrate the spherical law of sines that tells that $\frac{\sin(u_{ij})}{\sin(p_{ij})} = k$, with $k$ independent of the indices, and we determine the value of $k$. This law will be useful to prove the formula that gives the axial part of the composition of the quaternions. We use Figure A1(b), which is the same as Figure A1(a), except that the vectors $\mathbf{u}_i$ are such that the norms of their cross-products are units, i.e. $\|\mathbf{u}_i \times \mathbf{u}_j\| = \|\mathbf{u}_i\|\|\mathbf{u}_j\|\sin(u_{jk}) = 1$. The norms of $\mathrm{u}_i$ should be such that $\mathrm{u}_i\mathrm{u}_j \sin(u_{ij}) = 1$, which leads to three equations that are easily solved. It comes that

$$\mathrm{u}_i = \frac{\sin(u_{jk})}{\sqrt{p}} \tag{A9}$$

$p = \sin(u_{12})\sin(u_{23})\sin(u_{31})$ and $\{i, j, k\} \in (1,2,3)$.

The vectors of the reciprocal lattice are $\mathbf{p}_i$ such that $\|\mathbf{p}_i\| = \frac{1}{h_i} = \frac{\|\mathbf{u}_j \times \mathbf{u}_k\|}{V} = \frac{1}{V}$. Similarly as the fact that the volume of the unit cell is given by $\det(\mathbf{u}_i, \mathbf{u}_j, \mathbf{u}_k) = \mathbf{u}_i \cdot (\mathbf{u}_j \times \mathbf{u}_k) = V$, with $\{i, j, k\} \in (1,2,3)$, we get for the reciprocal lattice $\det(\mathbf{p}_i, \mathbf{p}_j, \mathbf{p}_k) = \mathbf{p}_i \cdot (\mathbf{p}_j \times \mathbf{p}_k) = \frac{1}{V}$. In addition, the vector $\mathbf{p}_j \times \mathbf{p}_k$ is necessarily parallel to $\mathbf{u}_i$, thus we can write $\mathbf{p}_j \times \mathbf{p}_k = x\,\mathbf{u}_i$. The value of $x$ is such that $\mathbf{p}_i \cdot \mathbf{u}_i = 1$, which implies that $\mathbf{p}_i \cdot (\mathbf{p}_j \times \mathbf{p}_k) = x = \frac{1}{V}$. Consequently, $\|\mathbf{p}_j \times \mathbf{p}_k\| = \frac{\sin(p_{jk})}{V^2} = \frac{\mathrm{u}_i}{V}$. It comes that

$$\sin(p_{jk}) = V\,\mathrm{u}_i \tag{A10}$$

Combining equations (A9) and (A10) gives the spherical law of sines

$$\frac{\sin(p_{jk})}{\sin(u_{jk})} = \frac{V}{\sqrt{p}} \tag{A11}$$